\documentclass[]{mn2e}
\usepackage{graphics}
\title[Water ice formation]{On water ice formation in interstellar clouds}

\author[R. Papoular]{R. Papoular$^{1}$\thanks{E-mail:
papoular@wanadoo.fr}\\
$^{1}$Service d'Astrophysique and Service de Chimie Moleculaire,\\
CEA Saclay, 91191 Gif-s-Yvette, France}
\begin{document}

\date{Accepted . Received ; in original form }

\pagerange{\pageref{firstpage}--\pageref{lastpage}} \pubyear{2002}
   \maketitle
\label{firstpage}
\begin{abstract}
A model is proposed for the formation of water ice mantles on grains in interstellar clouds. This occurs by direct accretion of monomers from the gas, be they formed by gas or surface reactions. The formation of the first monolayer requires a minimum extinction of interstellar radiation, sufficient to lower the grain temperature to the point where thermal evaporation of monomers is just offset by monomer accretion from the gas. This threshold is mainly determined by the adsorption energy of water molecules on the grain material; for hydrocarbon material, chemical simulation places this energy between 0.5 and 2 kcal/mole, which sets the (``true") visible exctinction threshold at a few magnitudes. However, realistic distributions of matter in a cloud will usually add to this an unrelated amount of cloud core extinction, which can explain the large dispersion of observed (``apparent") thresholds. Once the threshold is crossed, all available water molecules in the gas are quickly adsorbed, because the grain cools down and the adsorption energy on ice is higher than on bare grain. The relative thickness of the mantle, and, hence, the slope of $\tau_{3}(A_{v})$ depend only on the available water vapour, which is a small fraction of the oxygen abundance. Chemical simulation was also used to determine the adsorption sites and energies of O and OH on hydrocarbons, and study the dynamics of formation of water molecules by surface reactions with gaseous H atoms, as well as their chances to stick in situ.

  \end{abstract}

%
\section{Introduction}

The present work addresses three questions:

1) Why does ice form in clouds only when protected from interstellar light by a significant amount of extinction ?

2) Why is the optical depth of the ice feature at 3 $\mu$, linearly related to extinction beyond formation threshold ?

3) Is it possible to quantitatively predict this behaviour on the basis of known physics ? 

These questions were addressed more than once in the past (see Whittet et al. \cite{whit01}, Whittet \cite{whit02}). However the answers given did not always concur. They went roughly along two different lines. Initially, van de Hulst \cite{vdh49} proposed that ice formation occurred upon chemical combination of O and H atoms on grain surfaces. Later, Jones and Williams  \cite{jon84}, for instance, developed this model based on educated guesses of sticking coefficients of atoms on grains.

In the meantime, it was also suggested that ice formed directly by water vapour deposition. But this model was countered by theoretical models of gas chemistry in clouds (Herbst and Leung \cite{herb86}), which pointed to low concentrations ($\sim2\ 10^{-8})$ of water vapour in clouds of interest. However, addressing the threshold problem, Williams et al. \cite{wil92} later did not exclude the possibility that this direct deposition indeed contributed. Besides, it should be clear that this scenario does not in any way imply that all water molecules are formed in the primordial gas.

The present work started in an effort to confirm the first condensation scenario by ascertaining quantitatively the estimates which were previously made of the various surface parameters involved. Along the way, it was found that water formed on a surface could not stick to it, because of its very large initial energy content ($\sim$5 eV). This made it difficult to pursue the first scenario. Besides, it was also found that depositing the energy of one 3-$\mu$ photon ($\sim$0.4 eV) in an adsorbed water molecule, as proposed by Williams et al. \cite{wil92}, is not enough to exctract it from the grain surface, making it necessary to seek another explanation for the threshold extinction. Hence the elaboration of a new model, based on the second scenario.

In the following, we reverse this chronological order, and start with a detailed description of the proposed model (Sec. 2), a cursory description of which follows. The first layer of ice-to-be develops when the abundance of water molecules in the ambient gas exceeds the vapour pressure corresponding to the grain temperature (supercooling). The latter, $T$, is determined by equating radiative powers absorbed and emitted by the grain: it is, therefore, dependent on the visual magnitude, $A_{v}$, characteristic of the extinction which shields cloud grains from outside radiation. As to the vapour pressure of water at this temperature, it depends essentially on $T$ and on the heat of adsorption, $E_{0}$, of a water molecule on the grain surface considered. $A_{v}$ at threshold is thus indirectly linked to $E_{0}$.

Once the first layer is completed, further mantle growth is limited only by the total water vapour available in the ambient gas (Sec. 3). This is because, as ice is deposited, the grain temperature falls while $E_{0}$ rises, so that sublimation is exponentially suppressed. For a given grain radius, the optical depth , $\tau_{3}$, at the peak of the ice feature, is then shown to be proportional to the increase of extinction over its threshold value. A relation is given between the corresponding coefficient of proportionality and the relative mantle thickness.

Assessing quantitatively this scenario (Sec. 4) requires the knowledge of the optical and surface properties of grain core and ice mantle, as well as heats of adsorption of water molecules on core and mantle. Laboratory data relative to the latter is inadequate: most studies are dedicated to adsorption of atoms on pure and regular surfaces of technical interest, at temperatures and pressures much higher than relevant here. However, the availability of new, powerful, software for chemical modelling has opened other avenues from which to seek the required answers. Adsorption (see Steele \cite{ste02}, Al-Halabi et al. \cite{alh04}), surface dynamics (see Kolesnikov et al. \cite{kol04}, Hasnaoui et al. \cite{has05}), crystal growth (see Henkelman et al. \cite{hen03})and chemical reactions between gas and solid surfaces (see Sadlej et al. \cite{sad95}) are now abundantly treated with such software.

The present work also resorts to numerical modeling to find adsorption sites and energies for H, O and H$_{2}$O on aromatic and aliphatic hydrocarbon surfaces typical of C-rich grains. From this data, and using the condensation model proposed above, a threshold extinction can be deduced which is in rough agreemant with observations.

\section{Deposition of the first H$_{2}$O layer}

Consider a molecule adsorbed on a uniform flat surface of a different material, at temperature $T$. The probability per unit time for it to desorb is

$p=\nu exp(-E_{0}/k_{B}T)$     (s$^{-1}$)

where $\nu$ is the attempt frequency, usually taken to be the vibrational frequency of the molecule in its potential well, $E_{0}$ its binding energy or potential depth (also enthalpy of sorption on, or activation energy of desorption from, the core surface) and $k_{B}$, Boltzmann constant (see Atkins \cite{atk98}, Somorjai \cite{som94}). Assume, ideally, that there are $s$ identical adsorption sites per unit area of the surface, a fraction $\theta$ of which is occupied by one adsorbate molecule each. The desorption rate is then given by

$R=s\theta\nu exp(-E_{0}/k_{B}T)$    (mol$\,$cm$^{-2}$$\,$s$^{-1}$)

If the molecule number density in the atmosphere above is $fn_{H}$, where $f$ is the molecule relative abundance and $n_{H}$, the atomic hydrogen density, and if we neglect for the moment the probability of a newly arriving molecule being adsorbed over a previously adsorbed molecule (see Discussion below), then the deposition or accretion rate is $fn_{H}V(1-\theta)/3$, where $V$ is the velocity of the gas molecules, and the condition for dynamical equilibrium at coverage $\theta$ of the surface is
\begin{equation}
1/3 fn_{H}V(1-\theta)=s\theta\nu exp(-E_{0}/k_{B}T).
\end{equation}

The mantle formation can be considered to have started when $\theta\sim0.5$. However, for the first monolayer to be completed in a reasonably short period, the desorption rate should be, at most, say, one tenth of the deposition rate. By definition, this sets the supersaturation ratio at 10. The threshold temperature is therefore obtained from eq.(1) by setting $\theta=0.5$ after multiplying the l.h.s. by 1/10:
\begin{equation}
  T_{thr}=\frac{E_{0}}{R_{B}K_{av}},
\end{equation}
where $R_{B}$=1.98 cal/mole\,$\char'27$K and $K_{av}=ln(30s\nu /fn_{H}V)$, where the subscript ``av" stands for ``available" density of gas molecules. For the relevant values of $E_{0}$, and because of the exponential, $T$ is only very weakly sensitive to relatively strong variations in the components of $K_{av}$. In zeroth approximation, therefore, the poorely known actual abundance, $f$, can be replaced by the total oxygen abundance in the gas, although the available amount for ice condensation is known to be much smaller, because of oxygen depletion in CO, OH and grains. $E_{0}$, $s$ and $\nu$ depend on the nature of the grain surface and will be estimated in Sec. 4.

Now, the temperature $T$ of a grain inside a molecular cloud is essentially determined by the amount of radiative energy falling upon the grain, and this in turn depends on the extinction which screens the grain from the outside radiation. Both relations pave the way to the threshold extinction for the onset of ice formation.

Let the radiative flux at the grain location be $F=GF_{0}$, where $F_{0}$ is the interstellar radiation field ($10^{-9}$ W\,cm$^{-2}$) and $G$, the extiction coefficient, which is approximated by $G=exp(-A_{v}/1.08)$ (see Mezger et al. \cite{mez82}).

While the grain, supposed to be spherical, of radius $a$, is still bare, its equilibrium temperature is determined by equating absorbed and emitted radiative powers or, approximately
\begin{equation}
GF_{0}\pi a^{2}\alpha_{v} 2a=4\pi a^{2}2\alpha_{ir}a\sigma_{SB}T^{4},
\end{equation}
where $\sigma_{SB}$ is Stefan-Boltzmann constant ($5.7 \ 10^{-5}$ erg\,cm$^{-2}\char'27$K$^{-4}$); $\alpha_{v}$ and $\alpha_{ir}$ are,respectively, the absorption coefficient of the grain material in the visible range and in the infrared range corresponding to grain emission at temperature $T$. This approximation assumes that the optical depth of the grain is that of a cube of side $2a$, noting that both sides of the equation differ from this by the same factor. Now, $\alpha_{ir}$ can be approximated by its value at the wavelength of peak black-body emission at $T$, i.e. $\lambda_{max}=3000/T$ (in $\mu$ and deg K). Moreover, it is generally found that, in the far infrared, which is the emission range of interest here, $\alpha_{ir}\propto 1/\lambda$; that is also the case in the interstellar medium (see Dupac et al. \cite{dup03}). Thus, within the present degree of approximation,

$\alpha_{ir}\sim\frac{\alpha_{0}\lambda_{0}}{\lambda_{max}}=\frac{\alpha_{0}\lambda_{0}T}{3000}$,

where the subscript $0$ refers to an arbitrary wavelength within the infrared emission range of interest. As a result, eq.(3) reduces to 
\begin{equation}
T=CG^{1/5}$, with $C=10(\frac{1.3\alpha_{v}}{\alpha_{0}\lambda_{0}})^{1/5}.
\end{equation}
Using the definition of $G$ above,

$T=Cexp(-\frac{A_{v}}{5.4})$.

Reasonable values for the grain optical properties are, for instance,

$\alpha_{v}=2 \ 10^{4}$ cm$^{-1}$, $\lambda_{0}=10 \mu$, $\alpha_{0}=100$ cm$^{-1}$.

With these values inserted, eq.(4) gives $C=19$, so that 
\begin{equation}
T=19 exp(\frac{A_{v}}{5.4}),
\end{equation}
a relation plotted in Fig. 1. The latter also shows the impact of uncertainty in the optical constants.

\begin{figure}
\resizebox{\hsize}{!}{\includegraphics{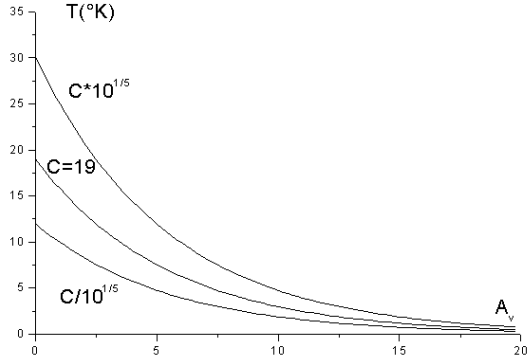}}
\caption[]{Adopted relation between the temperature $T$ of a grain and the visual extinction, $A_{v}$, in front of the grain location in the cloud; for 3 sets of values of the grain optical constants, in eq. 4, corresponding to C=19/10$^{-1/5}$ (lower curve), 19 (intermediate) and 19 10$^{1/5}$ (upper).} 
\end{figure}

Finally, substituting in eq.(2), the threshold extinction is
\begin{equation}
A_{v,thr}$=12.5 log$_{10}$($\frac{RCK_{av}}{E_{0}}).
\end{equation}
The physics of the condensation process is contained in $K_{av}$ and $E_{0}$. These depend on the particulars of the interactions of water molecules with the given dust, and are not directly available from the literature. Their numerical estimation will be taken up in Sec. 4. Here, in order to get a feeling of the relevance of this model, we shall study the dependence of $A_{v,thr}$ on $E_{0}$, considered as a free variable, and on $K_{av}$, considered as a parameter determined for a few sets of values of the physical quantities entering in eq.(6).

Since there are 5 such quantities, we chose a reasonable median average value for each and then add to $K_{av}$ a number of times $\pm$ln10, thus allowing for large variations of any of the 5 quantities. For the median values, we take

$s=10^{15}, \nu=5 \,10^{11}, n_{H}=10^{3}, f=7\ 10^{-5}$,

to get $K_{av,median}=58.3$ and, hence,with $C=19$, Fig. 2. Clearly, the particular choice of these values has no great impact on the behaviour of $A_{v,thr}$ as a function of the adsorption energy. More weight is carried by the optical parameters entering in the constant C (eq. 4), although this is mitigated by the exponent 1/5: Fig. 3 shows the effect of multiplying the median C by $10^{-1/5}$ or $10^{1/5}$.

\begin{figure}
\resizebox{\hsize}{!}{\includegraphics{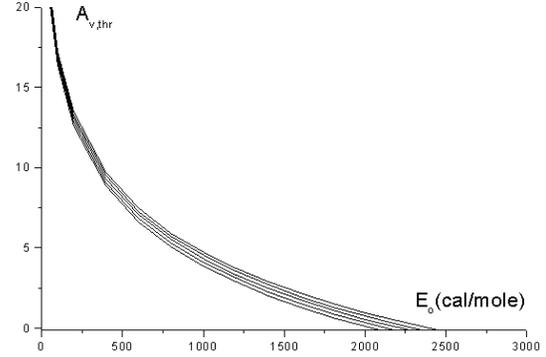}}
\caption[]{The threshold visual extinction as a function of the adsorption energy, according to eq.(6), for 5 sets of values of adsorption parameters, corresponding respectively to $K_{av}=K_{av,median}+Nln10$, with N= -2, -1, 0, 1 and 2, from bottom to top; N=0 corresponds to the median values  defined  after eq.(6). In all cases, the optical parameters are the same: $C=19$.}
\end{figure}

\begin{figure}
\resizebox{\hsize}{!}{\includegraphics{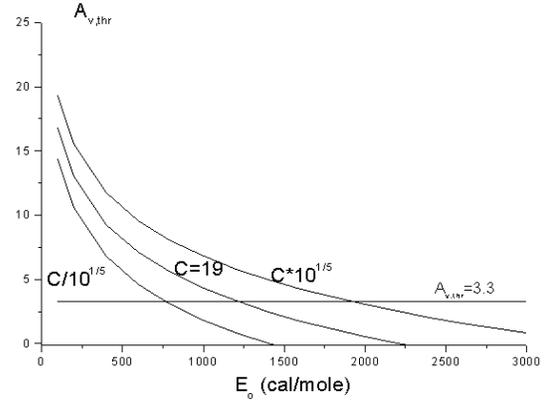}}
\caption[]{The threshold visual extinction as a function of the adsorption energy, according to eq.(6), with the same, median, adsorption parameters for the 3 curves, but different optical parmeters: $C=19*10^{-1/5}$, 19 and $19*10^{1/5}$, from bottom to top.} 
\end{figure}

The horizontal line $A_{v,thr}=3$, in Fig. 3, corresponds to the Taurus cloud as observed by Whittet et al. \cite{whit88}. Obviously, such a value can be predicted by the present model, on the basis of reasonable or available values of the parameters involved.

However, a wide range of threshold values , from 3 to 30, has also been observed (see Williams et al. \cite{wil92}), which prompted the latter authors to propose a model based on water desorption upon absorption of resonant infrared photons (3 $\mu$) from the iinterstellar medium. This is discussed in Sec. 4, but, in the framework of the present model, a different explanation is required. This can be sought in the definition of the threshold. In the above, it was tacitly assumed that the density distribution within the cloud was single peaked and asymmetric (like curve a, Fig. 4), with $A_{v}$ corresponding to the matter between 0 and $L/2$, distances being counted from the star, along the sightline, and ice first appearing at the peak, at $L/2$. It seems much more realistic to envision a nearly symmetric distribution (Fig. 4b), in which case, the ``apparent" threshold is $2A_{v,thr}$. Even this cannot always be the case: more probably (see Ward-Thompson  et al. \cite{war03}), the matter in the inner cloud will contribute, in the eye of the observer, additional extinction which bears no relation to $2A_{v,thr}$, which is contibuted by the two limbs (Fig. 4c). We therefore expect $A_{v,thr}$ to set only a lower limit to the measured, ``apparent" threshold.

\begin{figure}
\resizebox{\hsize}{!}{\includegraphics{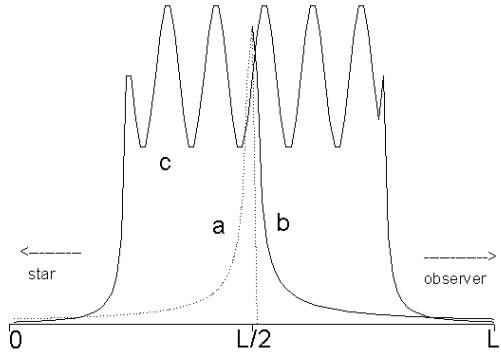}}
\caption[]{Possible schematic distributions of optically absorbing matter along a sight line through a molecular cloud.} 
\end{figure}

Although the other, incompletely known, parameters certainly contribute some dispersion of observed thresholds, they can hardly explain a wide dispersion among clouds which are supposed to harbour a ubiquitous mixture of known dust types.

\section{Steady-state ice mantle}

Whittet et al. \cite{whit88} showed that, along various lines of sight through the Taurus cloud, the optical depth at the peak of the ice feature, $\tau_{3}$, was a linear function of the corresponding extinction:

$\tau_{3}=q(A_{v}-A_{v,thr})$,  q=0.072$\pm$0.002.

Such a regular relation seems to imply that a) the grains have accreted a constant relative amount of ice, b) this amount is the same throughout the cloud. This strongly suggests that all water molecules formed through gaseous or surface reactions in the cloud are quickly deposited on the grains in the inner cloud, which are shielded by $A_{v,thr}$ or a still higher extinction, and are therefore cooler still. For a given cloud, the number of available water molecules is essentially the complement of the number of oxygen atoms which are included in CO, OH and grains; it is thus a fraction of the O abundance. Let us estimate this fraction on the basis of observations. For grains of radius $a$, uniformly covered by a mantle of thickness $e$, the abundance of available water molecules, relative to that of H atoms, assumed all to be deposited in the mantle, is 
\begin{equation}
f=\frac{4\pi(a+e)^{2}ed_{ice}n_{gr}}{n_{H_{2}O}n_{H}}
 = \frac{1}{6}\frac{M_{d}}{M_{H}}\frac{d_{ice}}{d_{gr}}\frac{e}{a}(1+\frac{e}{a})^{2},
\end{equation}
where $M_{d}/M_{H}$ is the dust-to-gas mass ratio, $d_{ice}$ and $d_{gr}$ the mass densities of ice and grain materials respectively.

On the other hand, for the same geometry and taking $N_{H}= 2 \ 10^{21} \Delta A_{v}$, where $\Delta A{v}=A_{v}-A_{v,thr}$,
\begin{equation}
\frac{\tau_{3}}{\Delta A_{v}}=\frac{\pi(a+e)^{2}.2e\alpha_{ice}.n_{gr}.L}{n_{H}L/2 10^{21}}
=5 \ 10^{-3}\frac{M_{d}}{M_{H}}\frac{\alpha_{ice}}{d_{gr}}\frac{e}{a}(1+\frac{e}{a})^{2}.
\end{equation}
Combining the last two relations and taking $M_{d}/M_{H}=1/150$, $d_{ice}=1$ and $\alpha_{ice}=2.6\  10^{4}$ cm$^{-1}$ (Leger et al. \cite{leg83}), one finally obtains
\begin{equation}
f=1.3\ 10^{-3}\frac{\tau_{3}}{A_{v}}.
\end{equation}
For q=0.072 (Taurus cloud, see above), $f=9.4\ 10^{-5}$, comfortably smaller than the total O abundance. More generally, this relation can be used for a better approximation than that which was adopted to plot Fig. 2.

Eq. (8) also gives $e/a$ as a function of q, a relation plotted in Fig. 5, with 
$\alpha_{ice}=2.6\ 10^{4}$cm$^{-1}$ and $d_{gr}=2$. For $q=0.072$, $e/a=0.65$.

\begin{figure}
\resizebox{\hsize}{!}{\includegraphics{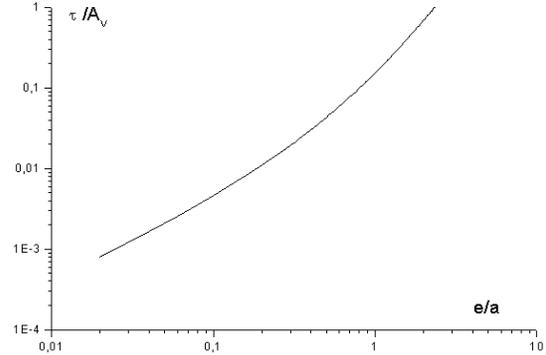}}
\caption[]{The relative mantle thickness $e/a$  as a function of $\tau_{3}/\Delta A_{v}$).} 
\end{figure}

The assumption that all available water molecules are promptly deposited as a mantle on grains in the cloud core requires some justification in view of the possible obstacle of surface tension. The tension over the curved surface of the mantle generates inside an additional pressure which enhances sublimation from the surface. This effect is greatest at the onset of mantle formation, when the outer radius is smallest, and may thus stop deposition altogether if the core radius is too small. The critical radius for this to occur is deduced from Lord Kelvin's expression for the ratio, $S_{K}$, of water vapour pressure over a sherical grain of ice to that over a plane of the same material. If partial vapour pressures are expressed in terms of hydrogen density,

$S_{K}(ice)=\frac{f(T,a,ice)}{f(t,\infty,ice)}=exp\frac{2\gamma(ice)m_{H_{2}O}}{d(ice)k_{B}Ta}$,

where $\gamma(ice)$ is the surface tension, or capillary constant, of ice. By definition, this is the energy required to create a unit surface area of free ice surface, by breaking all bonds along a plane drawn through the bulk material. This can be approximated by $\gamma(ice)=sE_{0}(ice)/2$, where $E_{0}(ice)$ is the adsorption energy of the vapour monomer on the ice surface (which is close to the bond energy between monomers in bulk ice) , and $s$, the surface density of adsorption sites. Whence,
\begin{equation}
S_{K}(ice)=exp(\frac{sm_{H_{2}O}}{d(ice)}\frac{E_{0}(ice;cal/mole)}{k_{B}aT}).
\end{equation}

Now, the first fraction on the r.h.s. is nothing but the ratio of surface to volume densities of water monomers in ice, i.e., the average distance, $\delta$, between monomers, assumed to be the same on the surface and in bulk. This gives finally

$S_{K}(ice)=0.5\frac{\delta}{a}\frac{E_{0}(ice)}{T}$

where $E_{0}$ is in cal/mole. The mantle will grow only if the actual pressure of water vapour in the atmosphere, $f$, exceeds $f(T,a,ice)$. Now, in Sec. 2, we defined threshold by the condition

$f=10 f(T,E_{0}(ads))$,

where ``ads"  serves only as a reminder that the bond energy to be used here is that of a monomer on a bare grain surface as opposed to $E_{0}(ice)$, which characterizes bulk ice. The condition for the ice mantle to grow can now be written

$10f(T,E_{0}(ads))\geq S_{k}(ice)f(T,\infty,ice)$,

where the $f$'s are given by eq.(1) and depend on the respective $E_{0}$'s but not on $a$. Taking logarithms of both sides,

$2.3+ \frac{(E_{0}(ice)-E_{0}(ads)}{2T} \geq  \frac{\delta}{2a}(\frac{E_{0}(ice)}{T}$.

The adsorption bond energy will be shown below (Sec. 4) to range between 0.5 and 2 kcal/mole, while $E_{0}(ice)$ is at least 5 times larger (because of the strong polarity of water), so that the term 2.3 can be neglected in comparison. Then the energies in the other terms approximately factor out, so that the condition for mantle growth  reduces roughly to 

             $a\geq\delta$.

This growth condition is trivial, and made so because the adsorption energy of the monomer on the bare grain is so much smaller than the bond energy in the bulk, so that the condition for first layer deposition automatically satisfies the surface tension constraint. 

It is also necessary to check that the time required for mantle deposition is only a fraction of  the cloud age. From eq.(1), Sec. 2, the time for deposition of a monolayer is

$t_{1L}\sim\frac{3s}{fn_{H}V}$   (s).

In the Taurus case, $f$ was found above to be $\sim10^{-4}$. Assuming $n_{H}=10^{3}$ cm$^{-3}$ and $T=10$ K at threshold, $t_{1L}$ is found to be $\sim10^{5}$ y.

As for the complete mantle deposition time, note that the water vapour flux at threshold is $fn_{H}V/3$; on the other hand, in general, the bare grain optical cross-section is $\sim10^{-21}$ cm$^{2}$, so the grain surface available for deposition is $\sigma\sim4\ 10^{-21}$cm$^{2}$
per free hydrogen atom (Spitzer \cite{spi78}). Thus, the total amount of water ultimately deposited as ice is $fn_{H}V\sigma/3$ per second per H atom, while the total number of water molecules assumed to be available in the atmosphere is $f$ per H atom. The time required for deposition is, therefore,

$t_{mantle}=\frac{3}{n_{H}V\sigma}$   (s).

For the same $n_{H}$ and $t$ as assumed above, this yields $\sim3\ 10^{6}$ y. Note that this time is independent of grain radius, mantle thickness or water vapour abundance in the atmosphere.

Thus, both deposition times are reasonably short enough for the model to apply. This model can be used for predictive purposes. In the particular case of the Taurus cloud, for instance, the observations of Whittet at al. \cite{whit88} suggest that the ``true" threshold is, at most, 3.3 mag. Figure 3, here, indicates that this range corresponds roughly to $1000\leq E_{0}\leq2000$ cal/mole, which may help characterize the dust on which ice condensed. Also, since no ice is positively observed in the diffuse interstellar medium, where $A_{v}=0$, the same figure predicts that no dust material has $E_{0}\geq3$ kcal/mole. Vice versa, if that dust is fully characterized beforehand, Fig. 3 will predict the ``true" threshold for a given  $E_{0}$. Either way, prediction requires prior knowledge of the adsorption bond energy of water on the candidate dust material. Unfortunately, very little seems to have been published so far on the materials of interest to astronomy, which brings us to the subject of Sec. 4.

\section{Atomic, molecular and surface properties}

The discussion above emphasized the need for quantitative data on which to base and justify a model. Relevant data in the literature are meagre at present, most of the recent effort being directed to interactions of atoms with pure, regular surfaces of materials of technical interest (see Somorjai \cite{som94}, Bortolani et al. \cite{bort94}, Masel \cite{mas96}): water and ice are certainly not a priority even though an impressive specialized review of this case was given some time ago by Thiel and Madey \cite{thi87} . Masel points to the odd behaviour of H$_{2}$O and CO, which do not follow  general trends; he gives typical physisorption energies as 2 to 10 kcal/mole. Dormant and Adamson \cite{dor68} studied the adsorption of nitrogen on molecular solids. For ice, in particular, they give $E_{0}$ between 1500 and 2500 cal/mole.

  More to our point, Zettlemoyer et al. \cite{zet75} studied the adsorption of water on several solid surfaces, including  carbon blacks, for which they find adsorption energies $\sim10$ kcal/mole. But Suzuki and Churchill \cite{suz90} cited a range of 1 to 10 kcal/mole for common gases on sieving carbon. Ibupoto and Woods \cite{ibu01} measured about 600 cal/mole for water on cellulose. 

These widely dispersed results were also generally obtained at about 100 K and relatively high pressures. They can hardly be extrapolated to the special cases of adsorption of O, OH and H$_{2}$O on amorphous ice and hydrocarbons at much low temperature and pressure (see further discussion in Sec. 5).
 
Of course, new experiments will ultimately give the answers, but numerical chemical simulation is now able to deliver more quickly an estimate of the required quantities relevant to the special case at hand. It is used here, instead.

\subsection{The chemical code}
A description of the software package used for simulation can be found in Papoular \cite{pap01}. The particular code used here is AM1 (Austin Model 1), as proposed, and optimized for carbon-rich molecules, by Dewar et al. \cite{dew85}. This semi-empirical method combines a rigorous quantum-mechanical formalism with empirical parameters obtained from comparison with experimental results. It computes approximate solutions of Schroedinger's equation, and uses experimental data only when the Q.M. calculations are too difficult or too lengthy. This makes it more accurate than poor ab initio methods, and faster than any of the latter.

The code is used in the first place to create a plausible model grain: once the size, composition and general structure have been chosen, the ``optimization" procedure determines the correct relative positions of the atoms by minimizing the total potential energy. In order to study the interaction of a given species (atom or molecule) with the grain surface or with species previously adsorbed on it, the new species is created at some distance from the grain and given some velocity towards it. The molecular dynamics is then launched.

The molecular dynamics relies on the Born-Oppenheimer approximation to determine the motions of atoms under nuclear and electronic forces due to their environment. At every step, all the system parameters are memorized as snapshots so that, after completion of the run, a movie of the reactions can be viewed on the screen, and the trajectory of any atom followed all along. This makes for a better understanding of the details of  mechanisms and outcomes. Note that the dynamics of bond dissociations and formation can only be simulated by using Unrestricted Hartree-Fock (UHF) wave functions in the Q.M. part of the calculation (see Szabo and Ostlund  \cite{sza89}). The elementary calculation step was set at $10^{-15}$ s. 

Semi-empirical simulation methods do not account for purely quantum-mechanical phenomena like tunneling or zero-point energy. In the present context of relatively high grain temperature, the former is negligible as regards surface diffusion, and the latter hardly affects the probability and energetics of the physico-chemical reactions involved.
\begin{figure}
\resizebox{\hsize}{!}{\includegraphics{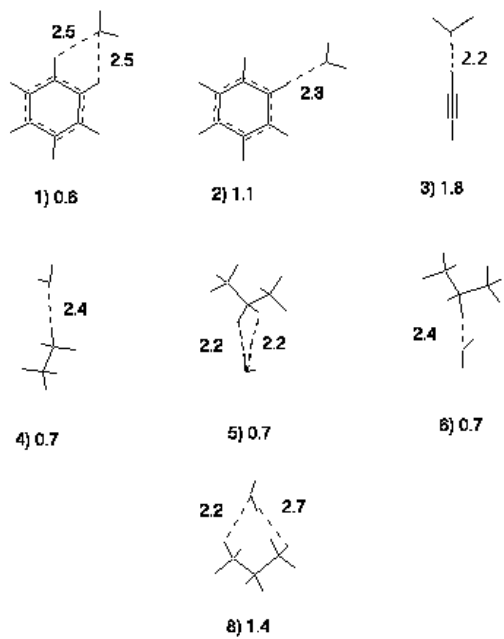}}
\caption[]{Main adsorption sites and energies near typical hydrocarbon surface functional groups. Aromatic CH groups are recognized, as well as aliphatic CH, CH$_{2}$ and CH$_{3}$ groups, in the conventional representation. In each case, the independent V-shaped molecule is water, trapped in the adsorption site, and its oxygen atom is linked to the nearest H atom or atoms on the ``grain" by dotted lines whose lengths are indicated by near-by numbers in angstroms. Below each system appears its ranking number, followed by the corresponding adsorption energy, $E_{0}$, in kcal/mole.  One typical site not represented in this figure is above the center of aromatic rings, at $\sim3.5\ \AA{\ }$ from the 6 carbon atoms, with the lowest adsorption energy: 0.5\ kcal/mole. (1\ kcal/mole=43.5\ meV)}
\end{figure}
The procedure just described was used by Papoular \cite{pap05} to study the interactions of atomic H with hydrocarbon grains. In the course of that work, the adsorption potential wells of the model grains were determined in position and depth. The present work made use of these results as a guide for the adsorption of O, OH and H$_{2}$O.

\subsection{Adsorption sites and energies of H$_{2}$O}
This study is restricted to carbonaceous IS (InterStellar) dust. IR spectra of such dust reveals the presence of both aromatic and aliphatic C-C and C-H bonds. In order to save computation time, we have modeled several small structures presenting the most typical functional groups. Adsorption sites can be found by tossing the water molecule towards one of these targets, with some initial velocity. If this velocity is small enough, the projectile will dwell on the target periphery and wander over it, progressively loosing energy to the target, until it settles at the bottom of one of the possible adsorption wells created by the spatial variations of the van de Waals potential of  the structure.

 A quicker procedure is based on the use of the previous results on H adsorption: since the sites roughly coincide, the water molecule is initially positioned in one of the known sites of H adsorption, all velocities being set to zero, and the potential optimization of absorbant and adsorbate together is then launched; note the (possibly) new position of the adsorbate, and let $E_{min}$ be the minimum potential thus found; now, extract the water molecule and read the new potential of the structure, say $E_{grain}$. Then, the adsorption energy can be deduced to be 

$E_{0}=E_{grain}+E_{H_{2}O}-E_{min}$,

 where $E_{H_{2}O}$ is the internal potential energy of a water molecule, and is computed by the code to be 223 kcal/mole. $E_{grain}$ usually does not differ much from the potential energy of the target alone, because van der Waals forces are so weak relative to the chemical binding energies.

The initial temperature of each system can be controled through the chemical code. Here, in view of the very low temperatures known to be required for ice formation, the initial temperature was uniformly set at 0 K. When the water molecule is adsorbed, energy $E_{0}$ is shared between adsorbate and absorbant; because the latter is normally so much larger than the water molecule, the consequent rise in temperature is negligible and the adsorbat remains locked to the surface.

 Figure 6 summarizes the result and shows that the adsorption energies determined as described fall between 0.5 and 2 kcal/mole; according to Fig. 3, the proposed model is therefore coherent with observations. It also shows that, because these energies are so weak, ice can hardly form in the diffuse interstellar medium, and needs at least a few visible magnitudes of shielding. This result is discussed further in Sec. 5.

It is noteworthy that an average bonding energy between monomers in ice can be deduced from the latent heat of sublimation of ice in vacuum, 600 cal/g, to be 18*600*10$^{-3}=10.8$ kcal/mole, much larger than adsorption energy of a monomer on a bare grain, and nearer to the results of temperature programmed desorption (see Kay et al. \cite{kay89}, Fraser et al. \cite{fra01}). In order to understand this behaviour, successive layers of monomers were added on each other to form an ellipsoidal chunk of 27 molecules. By minimizing its potential energy at zero temperature, this chunk was ``quenched" into an amorphous state in which all monomers fit tightly, but randomly, together. Starting from the outer ones, the monomers were then extracted in turn, noting the new potential energy at each step. From this it was deduced that the binding energies ranged from 18 to 2 kcal/mole, or roughly 10 on average. Ice monomers hold together more tightly than to the substrate, because hydrogen bonds are stronger than van der Waals forces.

Further quantitative information can be gained from the chemical model used in this section. Thus, based on the average distance of adsorption sites from the outer atoms of the grain ( Papoular \cite{pap05}), we estimate the average distance between these sites to be about 3 $\AA{\ }$. This justifies setting $s=10^{15}$ cm$^{-2}$ for the surface density of states in eq. (1). Moreover, once a water molecule has settled at the bottom of one adsorption potential well, it can be displaced by, say, $\sim1  \AA{\ }$ and then left to oscillate about the bottom while monitoring the dynamics all along by means of the chemical software. This gives us the attempt frequency, $\nu$, used also in eq. (1), about $5\ 10^{11}$ s$^{-1}$, on average. This value falls in the range typical of (the relatively weak) physical bonds ,$\sim 10^{12}$ s$^{-1}$, as opposed to $\sim10^{14}$ s$^{-1}$, which applies to (the stiffer) chemical or even H-bonds ( see Atkins \cite{atk98}).

\subsection{Ice formation on grains}

Chemical simulation also helps discussing the possibility of formation of ice on grains by  accretion and conversion of O atoms into water, followed by combination with two successive gaseous H atoms impinging on the solid surface. For this purpose, the procedure used previously for H adsorption, and above for water adsorption, was applied to O atoms. Here, for a more realistic model, a larger structure (18 C and 28 H atoms), similar to that used by Papoular(\cite{pap05}), and including  the functional groups of Fig. 6, was used instead of the individual functional groups. Adsorption sites were found along CH functional groups at $\sim2.2  \AA{\ }$ from the H atom, with energy $\sim0.6$  kcal/mole. This is enough to keep the adsorbate on the surface at the low temperatures of interest, i.e. the sticking probability can be taken to be 1. Next, a free H atom was directed roughly towards an adsorbed O atom. After some wandering it reacted with the latter to form an OH radical which was immediately expelled, with a high velocity ($1.5\ 10^{5}$ cm\,s$^{-1}$) and strong internal oscillations (70 kcal/mole), both tapped from the high chemical energy released in the process (111 kcal/mole). Expulsion occured within 1 period of internal oscillation of the radical ($\sim10^{-14}$ s), which is much shorter than the relaxation time of the grain ($10^{-13}-10^{-12}$ s).

The same procedure was applied to the second step of this scenario, viz. accretion of a gaseous H atom by an adsorbed OH radical, to form water. Here, by contrast with OH formed in situ, the OH radical is first independently optimized in potential energy, and deposited with zero atomic velocities near an adsorption site. The whole system is then optimized to find the adsorption energy, $\sim1$ kcal/mole. As before, a free H atom was then directed towards the adsorbed OH radical, and ultimately formed with it a  new water molecule. The latter was expelled from the surface with high internal and translational energies, but, this time, only after $\sim2$ ps and a large number of internal oscillations; by then, the grain had relaxed. 

Both the expulsions of newly formed OH and H$_{2}$O can be understood by monitoring the motions of their atoms as well as of the nearest H atoms of the target grain. When a new OH bond is formed, the H atom oscillates with a large amplitude and, in so doing, encounters surface H atoms. At each encounter, part of its internal energy is lost to the grain and another part is transformed into external kinetic energy of the new species. When the latter becomes about equal to the adsorption energy of the species on the grain, the species escapes. Expulsion is favored by low adsorption energies and high formation energy of the new species. The grain size does not matter as long as it is much larger than the said species.

Thus, although water molecules may form on grains according to van de Hulst suggestion, neither the intermediate OH, nor the ultimate H$_{2}$O can stay in situ. Direct accretion of water from the gas is required to start ice mantle formation.

\subsection{Photodesorption of water}
Williams et al. \cite{wil92} proposed that, if enough interstellar radiation at 3 $\mu$ is allowed on adsorbed water molecules, they can be desorbed because this wavelength is resonant with their natural  frequency, and can therefore impart much energy to them; and this would explain the existence of a threshold for ice formation. In order to test this hypothesis, a water molecule was deposited with zero velocity in an adsorption site; one of its OH bonds was then extended from its equilibrium length (0.96 $\AA{\ }$), to 1.2 $\AA{\ }$, which gave the molecule an excess potential energy of 10 kcal/mole (the energy of a 3$\mu$ photon). Strong oscillations set in, but the molecule was not desorbed. The initial excess  energy was then increased by steps till the molecule was observed to desorb. This did not happen before the excess reached near  the OH bond energy (111 kcal/mole), in agreement with the finding above that OH or H$_{2}$O formed on the surface by combination of O or OH with a free H atom, is immediately desorbed. Thus, a single infrared photon can hardly desorb an adsorbed monomer. 

\section{Discussion}

Section 2 addresses only the very beginning of mantle formation, when water has not yet filled all the available adsorption sites on the bare grain surface. Equation 1 assumes that, in this sub-monolayer regime, the desorption rate is proportional to the surface coverage by water molecules. In general, this ``first-order" model is considered to be legitimate in as much as it is applied to the initial stage of adsorption, under very low temperature and vapour pressure, when the grain is nearly bare and the molecules hit it one at a time and stay put wherever they are adsorbed. Under such circumstances, before a large fraction of the grain surface has been covered with a monolayer of adsorbate, a newly arriving molecule stands only a minute chance of falling precisely on a previously adsorbed one, so that adsorption and desorption are essentially independent processes, both being, however, controled by interactions of a single adsorbate molecule with the bare surface. This is the basis of Langmuir's adsorption isotherms, which is found to apply most generally, especially to non-metallic substrates (see Masel \cite{mas96}, Atkins \cite{atk98}).
 
Later on, incoming molecules pile up upon one another, forming a multilayered structure. For water and alcohols, prone to H-bond formation, the measurements of desorption rates from such multilayers over plane substrates generally point to plane-like layers, from which desorption proceeds ``layer-by-layer", according to a so-called ``zero-order" rate (e.g. Kay et al. \cite{kay89}, Fraser et al. \cite{fra01}, for water ice); this is to mean that the desorption rate is now independent of coverage, unlike the sub-monolayer case. Cases have been observed where a fractional-order dependence is observed (see Nishimura et al. \cite{nis98}; Wu et al. \cite{wu93}, for alcohol ices). This multilayer regime is not relevant to the threshold problem but applies to ice mantle formation (Sec. 3): a non-zero desorption order (which does not seem to have been observed for water) could slow down the growth rate, but only slightly, for a given multilayer adsorption energy.

Nontheless, in order to understand published experimental results, it may be useful to dwell on multilayer structures, which are most often encountered in the water literature. In such structures, adsorbate-adsorbate interactions dominate over adsorbate-substrate bonding. As a consequence, both $E_{0}$ and the pre-exponential factor in the desorption rate may differ from those of the sub-monolayer regime. For hydrogen-rich species like water or alcohol ice, hydrogen bonds between adsorbates make for values of $E_{0}$ higher than those typical of the common van der Waals bonds responsible for physisorption. Thus, while $E_{0}$ is of order 10 kcal/mol for internal bonds of water and alcohol ices (see above and Fraser et al. \cite{fra01} and Nishimura et al. \cite{nis98}), it is reduced, for instance, to about 2 kcal/mol for liquid water, in which H-bonds do not form. 

In general, when the bond energies of the adsorbate to itself and to the substrate are distinct but slightly different, both can be deduced by analysis of the two corresponding peaks of desorption rate as a function of temperature, in the multilayer and sub-monolayer regimes, respectively (see Redhead \cite{red62}). For water on a gold substrate, the desorption curves yield only the water-water bonding characteristics, because they do not display a well-resolved sub-monolayer peak. Kay et al. \cite{kay89} interpreted this result as an indication that water binds more strongly to itself than to gold, to which it does not bond chemically. Zettlemoyer et al. \cite{zet75} also found that carbon blacks were hydrophobic. This hydrophobicity  strongly favours tight water-water bonding over weak water-substrate bonding, even in the sub-monolayer regime, the more so the higher the temperature and pressure. As a consequence, the adsorbed molecules would tend to bind together and segregate in 3-D islands, and desorption would, very early on, convert to a zero-order course, indistinguishable from the later course. This example shows the difficulty of determining experimentally physisorption bond energies, which requires detailed and accurate sub-monolayer measurements at very low temperature (see Masel \cite{mas96}). In a rare, but successful endeavour, Andersson et al. \cite{and84} studied the adsorption of water on Cu(100) and Pd(100) around 10 K, at submonolayer coverages. Using electron energy loss spectroscopy (EELS), they showed that water adsorbed as monomers with the oxygen end towards the substrate, and identified a hindered translational vibration at 28.5 and 41.5 meV, respectively. This corresponds to adsorption energies of 0.65 and 0.95 kcal/mol. They also observed the transition to clustered water beyond a temperature of 20 K or a coverage of 0.2.

Our chemical model for adsorption of water on a hydrocarbon substrate yields bond energies between 0.5 and 2 kcal/mol (depending on the particular site) and, correlatively, reveals no chemical or H-bonds between water and the model substrate (i.e. there is no electron sharing between water and substrate). From the known performance of the AM1 class of semiempirical models, the uncertainty in this estimate is a factor 2, similar to the intrinsic dispersion due to differences between adsorption sites.

Although it leaves to be desired, our estimate is qualitatively confirmed by the following facts:
\begin{itemize}
\item
the finding by Zettlemayer et al. \cite{zet75} that carbon blacks are hydrophobic, and by Kay et al. \cite{kay89} that water-substrate bonds are much weaker than internal water-ice bonds;
\item
the measurement of adsorption energies of water on Cu and Pd by Andersson et al. \cite{and84}, giving about 1 kcal/mol;
\item
the internal liquid water bond energy derived from its latent heat (80 cal/g) or surface tension (73 erg/cm$^{2}$) is $\sim2$ kcal/mol; CO-CO surface binding energy in pure CO ices is about the same (Sandford and Allamandola \cite{san88}); Al-Halabi et al. \cite{alh04} find a range of 1 to 3.5 kc/mol for CO on water ice;
\item
it is observed in the sky that mantle formation requires cooling inside shielding clouds down to about 10 K, while desorption of multilayer ice mantles in the laboratory occurs consistently above 130 K; if the internal (chemical) bond energy of water ice ($\sim10$ kcal/mol) were  to apply to our threshold model in Sec. 2 instead of the physisorption bond energy determined in Sec. 4 (0.5 to 2 kcal/mol), fig. 2 shows that $A_{v,thr}$ would always be $\sim0$, and ice would condense even in the diffuse ISM, neither of which is observed. 
\end{itemize}

A side effect of the high internal binding energy of water, the attendant formation of 3-D ``droplets" and the early setting of the zero-order regime, is that the threshold condition should no longer be defined by eq.(1) with $\theta\sim1$, but with a lower value. All other parameters being equal, this would decrease $K_{av}$ in eq. 2, increase the threshold temperature and hence, decrease the threshold extinction. For instance, Nishimura et al. \cite{nis98} for methanol on alumina, and Wu et al. \cite{wu93} for alcohols on NiO, observed the formation of ``droplets" as early as $\theta\sim0.5$. For a carbon substrate, Andersson et al. \cite{and84} find the transition at $\sim0.2$ monolayer. Even if $\theta$ were much smaller, that would not have dramatic effects on the threshold, thanks to the logarithmic expression of $K_{av}$ (eq.(6) and Fig. 2). As observed before, the desorption order in the following multilayer regime is not relevant to the threshold problem.

Among the other causes of uncertainty in our estimate of threshold extinction, a minor one is the value of the attempt frequency of water physisorbed on the model surface, $\nu=5 \,10^{11}$ s$^{-1}$ as compared with $5 \,10^{12}$ s$^{-1}$ for water H-bonded to its ice. Generally, for various materials, it is considered that typical values are  of order $10^{12}$ s$^{-1}$ for physisorption and $10^{14}$ s$^{-1}$ for chemisorption (see Atkins \cite{atk98}, Somorjai \cite{som94}, Masel \cite{mas96}), the strength of H-bonds being generally intermediate between those of physi- and chemisorption. Temperature desorption spectra yield $10^{13}$ s$^{-1}$ for methanol H-bonded on alumina (Nishimura et al. \cite{nis98}) and  $10^{14}$ s$^{-1}$ on NiO (Wu et al. \cite{wu93}). Our estimates fit well in this picture because a weaker bond results in a weaker restoring force and, hence, a lower oscillation frequency of the adsorbate in its potential well, i.e. a lower attempt frequency. Besides, even if the attempt frequency were in error by 1 or 2 orders of magnitude, the conclusions would not be affected notably, thanks again to the overwhelming weight of the exponential in eq.(1) (see fig. 2).

\section{Conclusion}
Chemical simulation supports a model of mantle formation by direct accretion of monomers from the gas, be they formed by gas or surface reactions. It also supports the threshold being defined by the minimum visible extinction of interstellar radiation which is necessary to lower the grain temperature to the point where thermal evaporation of monomers is just offset by monomer accretion from the gas.

 \section{Acknowledgments}
 
Thanks are due to an anonymous referee for fruitful comments on the manuscript.


\begin{thebibliography}{}
\bibitem[2004]{alh04}Al-Halabi A. et al. 2004, A\&A 422, 777
\bibitem[1984]{and84}Adersson S., Nyberg C. and Tengstal C. 1984, Chem. Phys. Lett. 104, 305
\bibitem[1998]{atk98}Atkins P. 1998, Physical chemistry, Freeman and Co., London
\bibitem[1994]{bort94}Bortolani V., March N. and Tosi M. 1994, Interaction of atoms and molecules with solid surfaces, Plenum Press, N.Y
\bibitem[1985]{dew85}Dewar M. and co-workers 1985, J. Amer. Chem. Soc. 107, 3902
\bibitem[1968]{dor68}Dormant L. and Adamson A. 1968, Journ. Coll. Interf. Sci. 28, 459
\bibitem[2003]{dup03}Dupac X. et al. 2003, arXiv:astro-ph/0310781
\bibitem[2001]{fra01}Fraser H., Collings. M., McCoustra M. and Williams D. 2001, MNRAS 327, 1165
\bibitem[1962]{hal62} Hall P. and Tompkins F. 1962, Trans. Farad. Soc. 58, 1734
\bibitem[2005]{has05}Hasnaoui A. et al. 2005, Surf. Sci. 579, 47
\bibitem[2003]{hen03}Henkelman G. et al. 2003, Phys. Rev. Lett. 90,116101
\bibitem[1986]{herb86}Herbst E. and Leung C. 1985, MNRAS 222, 689
\bibitem[1949]{vdh49}van de Hulst 1949, Rech. Astr. Obs. Utrecht 11, Part 2
\bibitem[2001]{ibu01}Ibupoto K. and Woods J. 2001, On-line Jour. Biol. Sc. 1(11), 1015
\bibitem[1984]{jon84}Jones A. and Williams. D. 1984,MNRAS 209, 955
\bibitem[1989]{kay89}Kay B., Lykke K., Creighton J. and Ward S. 1989, J. Chem. Phys. 91, 5120
\bibitem[2004]{kol04}Kolesnikov A. et al. 2004, Phys. Rev. Lett. 93,035503
\bibitem[1983]{leg83}Leger A. et al. 1983, A\&A 117, 164
\bibitem[1996]{mas96}Masel R. 1996, Pricinples of adsorption and reactions on solid surfaces,  J. Wiley \& sons, N.Y.
\bibitem[1982]{mez82}Mezger P. et al. 1982, A\&A 105, 372
\bibitem[1998]{nis98}Nishimura S., Gibbons R. and Tro N. 1998, J. Phys. Chem. B, 102, 6831
\bibitem[2001]{pap01}Papoular R. 2001, Spectrochemica Acta, part A, 57, 947
\bibitem[2005]{pap05}Papoular R. 2005, MNRAS 359(2),683
\bibitem[1962]{red62}Redhead P. 1962, Vacuum 12, 203
\bibitem[1995]{sad95}Sadlej J. et al. 1995, J. Chem. Phys. 102, 4804
\bibitem[1988]{san88}Sandford S. and Allamandola L. 1988, Icarus, 76, 201 
\bibitem[1994]{som94}Somorjai G. 1994, Introduction to surface chemistry and catalysis, J. Wiley \& sons, N.Y.
\bibitem[1978]{spi78}Spitzer L. Jr. 1978, in Physical processes in the interstellar medium, J. Wiley \& sons
\bibitem[2002]{ste02}Steele W. 2002, Appl. Surf. Sci. 196, 3
\bibitem[1990]{suz90} Suzuki M. and Curchill S. (edrs) 1990, Adsorption Engineering, Chemical Engineering Monograph 25, Elsevier, Amsterdam
\bibitem[1989]{sza89}Szabo A. and Ostlund N. 1989, Modern quantum chemistry, Mc Graw Hill, N.Y.
\bibitem[1987]{thi87}Thiel P. and Madey T. 1987, Surf. Sc. Rep. 7, 211
\bibitem[2003]{war03}Ward-Thompson et al. 2003, in Star formation at high angular resolution, ASP Conf. Ser., Vol S-221, edrs. Burton et al.
\bibitem[1988]{whit88} Whittet D. et al. 1988, MNRAS 233, 321
\bibitem[2001]{whit01}Whittet D. et al. 2001, ApJ 547, 872
\bibitem[2002]{whit02}Whittet D. 2002, Dust in the general environment, Institute of Physics, Bristol
\bibitem[1992]{wil92}Williams D. et al. 1992, MNRAS 258, 599
\bibitem[1993]{wu93}Wu M.-C., Truong C. and Goodman D. 1993, J. Phys. Chem. 97, 9425
\bibitem[1975]{zet75}Zettlemoyer  A., Micale F. and Klier K. 1975, in Water, a comprehensive treaty, ed. Franks F., Plenum Press, N.Y., chap5



\end{thebibliography}
\end{document}